\documentclass[10pt,aps,pra,twocolumn,showpacs,superscriptaddress,nobalancelastpage,notitlepage,longbibliography]{revtex4-2}

\usepackage{graphicx,color,mathbbol,amsthm,amsmath,amsfonts,amssymb,bm,braket,footmisc,multirow,latexsym,times}
\usepackage[normalem]{ulem}
\usepackage[colorlinks]{hyperref}
\hypersetup{
 colorlinks=true,
 citecolor=blue,
 linkcolor=blue,
 urlcolor=cyan
 }

\begin{document}

\title{Heisenberg scaling under collective non-parallel directional noise via geometric state design}

\author{Francisco Riberi}
\affiliation{\mbox{Electrical and Computer Engineering Department, University of New Mexico, 498 Terrace St NE, Albuquerque, NM 87106}} 
\affiliation{Center for Quantum Information and Control, University of New Mexico, NM 87131, USA}

\author{Lorenza Viola}
\affiliation{\mbox{Department of Physics and Astronomy, Dartmouth College, 6127 Wilder Laboratory, Hanover, New Hampshire 03755, USA}} 

\author{Milad Marvian} 
\affiliation{\mbox{Electrical and Computer Engineering Department, University of New Mexico, 498 Terrace St NE, Albuquerque, NM 87106}} 
\affiliation{Center for Quantum Information and Control, University of New Mexico, NM 87131, USA}

\begin{abstract} 
Spin-squeezed states enable entanglement-enhanced frequency estimation; however, the achievable performance is limited in practice by decoherence. We study the problem of parameter estimation under \emph{classical collective noise} that acts along a \emph{fixed} direction during signal encoding. While we restrict our analysis to Gaussian noise statistics, no assumption is made on the nature of the noise temporal correlations. Our approach captures both parallel (dephasing) and single-axis transverse noise as special cases, and covers both Markovian and non-Markovian metrological regimes. In the properly squeezed limit, where a Holstein-Primakoff description is accurate, we identify a geometric noise-immunity mechanism: for \emph{known} non-parallel signal and noise axes, an appropriate one-axis-twisted input encodes the signal in a quadrature that is metric-orthogonal to the direction of noise-induced diffusion. Irrespective of the noise temporal correlations, the resulting estimation precision exhibits Heisenberg scaling in the probe number. Optimal performance is achievable by measuring a single collective spin component. Imperfect knowledge of the noise axis produces a crossover from Heisenberg scaling at moderate probe number back to the known collective-dephasing bounds asymptotically.
\end{abstract}

\date{\today}

\maketitle

\emph{Introduction.} 
Entanglement-assisted metrology \cite{Giovannetti0,Giovannetti2,SmerziRMP,montenegro,huang2024} enables parameter estimation beyond classical limits. Precision measurements that leverage quantum effects have delivered experimental gains in applications as diverse as spectroscopy \cite{Wineland1,Leibfried2004,monika2010,Franke2023}, magnetic-field and amplitude-motion sensing \cite{Polzik,mussel2014,Gilmore2021,affolter2023}, time-keeping \cite{YeClocks,Colombo,robinson2024}, quantum imaging \cite{taylor2016,Catxere,aslam2023}, and gravitational-wave detection \cite{ligo2011,ligo2019,wang2023}. 
In a typical setting, $N$ non-interacting two-level atoms are used to sense a signal parameter $b$ with fixed resources. For uncorrelated probes, the error incurred by an estimator $\hat{b}$ obeys the standard quantum limit (SQL), $\Delta \hat{b} \propto N^{-1/2}$. This bound is not fundamental: in the absence of noise, maximally entangled Greenberger-Horne-Zeilinger (GHZ) states achieve Heisenberg scaling (HS), $\Delta \hat b \propto N^{-1}$, with even better precision emerging if probe interactions are allowed \cite{DattaPRL,riberi2025}. However, GHZ states are highly sensitive to decoherence and preparation errors \cite{SmirneReview}. Spin-squeezed states provide a scalable and experimentally more robust route to multipartite entanglement in atomic ensembles \cite{SchulteEchoes}, while still approaching HS when combined with suitable collective measurements \cite{Monika}.

In realistic settings, decoherence during the signal encoding limits the attainable metrological gain, with the resulting precision depending sensitively on both the noise geometry and its statistical properties. Here, we focus on \emph{directional} rank-one noise stemming from a \emph{classical}, stationary environment, modeled in terms of stochastic Gaussian fluctuations that couple to a \emph{collective} spin component along a fixed axis. Physically, collective noise tends to emerge due to probe proximity, and treating the noise as classical is justified for a variety of noise sources -- notably, stochastic fluctuations of the confinement fields and laser- or magnetic-field noise 
\cite{Degen,Rey2007,
Colombo}. Noise can be classified by its spatial correlations -- collective when all $N$ probes couple identically -- and temporal correlations -- white Markovian versus colored, depending on a zero or finite correlation time. A further key distinction is geometric: while dephasing, parallel noise  shares the signal's quantization axis, another important rank-one scenario arises when the noise acts along a distinct collective-spin direction.  

In the pure-dephasing limit, collective fluctuations are particularly detrimental: overcoming the SQL is impossible for both white and colored dephasing, as long as the noise is stationary and Gaussian \cite{MaNG}; the use of collective open-loop control does not lift this no-go \cite{third}. Known routes to overcome the SQL require additional resources, such as multiple transition frequencies enabling decoherence-free subspaces \cite{Dorner2012}, differential interferometry \cite{Landini2} or ancillary qubits \cite{AncillaBased}. By contrast, for dephasing noise that is spatially uncorrelated (local) or only partially correlated, super-classical scaling can be recovered in suitable regimes without such resources, to a degree that depends on the available noise knowledge \cite{FUR22,FUR}. 

When signal and noise act along non-commuting axes, improved precision scaling may be restored, depending on the underlying temporal correlations, by using pulsed dynamical decoupling \cite{Sekatski,DurDD,Lahcen}, continuous driving \cite{HaidongNature2,HaidongReview2}, 
or quantum error correction \cite{Sisi2018,Sisi2021,Mann,QECnote}. These approaches, however, demand fast control and measurement capabilities, and/or ancillary overhead, making them challenging to scale. The setting where noise is collective and non-parallel, and {\em no} extra resources are available remains scarcely  explored. For transverse directional noise with {\em no} spatial and temporal correlations, super-SQL scaling $\Delta \hat b \propto N^{-5/6}$ has been predicted for GHZ states by optimizing the duration of the encoding period \cite{Chaves2}. To our knowledge, no existing analysis addresses the collective noise limit without additional control or ancillas.

Here, we show that HS can be retained under \emph{collective} directional noise with a \emph{non-parallel} signal axis  and \emph{arbitrary} temporal correlations, provided the noise direction relative to the signal quantization axis is known. The key mechanism is geometric: by working in the Holstein-Primakoff regime, where the dynamics can be described in terms of Gaussian quadratures, the signal-induced displacement can be made metric-orthogonal to the noise-induced diffusion direction by tuning the ``shear'' in the one-axis-twisted input state. The quantum Fisher information then becomes insensitive to the noise, and the resulting performance is optimized by measuring a single collective spin component. We further quantify how uncertainty in the noise axis produces a crossover from HS at moderate $N$ to the known collective-dephasing asymptotic $(N\to\infty)$ bounds. 

\smallskip

\emph{Estimation setting.}
We consider an ensemble of $N \gg 1$ effective two-level systems for estimating an unknown static frequency, $b$, while subject to collective directional noise. We restrict the allowed initial states to the permutation-invariant sector with total spin $J \equiv N/2$ and collective spin operators $\mathbf{J}\equiv (J_x, J_y, J_z)$, where $J_i = \tfrac{1}{2} \sum_{n=1}^N \sigma_i^{(n)}$ in terms of spin-$1/2$ Pauli matrices. The noisy evolution is generated by a stochastic Hamiltonian of the form $H_{\rm S}(t)= b\, J_x +\xi(t) \,{\mathbf J}\cdot \hat{\mathbf{v}}$, where $|| \hat{\mathbf{v}}||=1$ and the real stochastic process $\xi(t)$ is assumed zero-mean, stationary, and Gaussian: $\langle \xi(t)\rangle_\xi=0$, for all $t$, and $\langle \xi(t)\xi(s)\rangle_\xi\equiv C(t-s)$, with $\langle \bullet\rangle_\xi$ and $C(\tau)=C(-\tau)$ denoting the ensemble average over noise realizations and the two-point noise (auto)correlation function, respectively. Noise is directional in the sense that one of the transverse components $\hat{\mathbf{v}}_\perp \equiv v_y \hat{\mathbf{y}} + v_z \hat{\mathbf{z}}$ is non-zero:
\begin{equation}
H_{\rm S}(t)= b\, J_x +  \xi(t) \big( J_x\cos\theta+ J_y\sin\theta\big),
\label{ham}
\end{equation}
where $\theta$ fixes the relative orientation between the signal and the noise axis in the $xy$ plane \footnote{The case $v_y=v_z=0$ ($\theta=0$) corresponds to pure dephasing, while $v_x=v_z=0$ ($\theta = \pi/2$) describes purely transverse noise. In writing Eq.\,\eqref{ham}, we assumed $v_z=0$. While the effect of a small $v_z$ component is addressed in the Supplement \cite{Suppl}, our approach would also apply by letting $v_y=0$, with appropriate modifications.   
}. 

We adopt a local estimation setting with a fixed total runtime $T$. Each run prepares an input state $\rho_0$, evolves it for a signal-encoding (or interrogation) time $t$, and measures an observable $\mathcal{O}$ on the time-evolved state $\rho(t)\equiv U_{\xi}(t)\rho_0 U_{\xi}^{\dagger}(t)$, where $U_\xi(t)$ denotes unitary evolution under the Hamiltonian in Eq.\,(\ref{ham}). While recent studies have considered noise correlations between successive experimental runs \cite{kaufmann2026}, here we assume that the the duration of each shot is sufficiently long  
for such correlations to decay and shot-to-shot fluctuations to be independent. 
Repeating the protocol $\nu\equiv T/t\gg 1$ times yields an estimator $\hat b$ near the operating point $b_0$. Performance is assessed via error propagation by the so-called method of moments \cite{SmerziRMP}, leading to the standard uncertainty \cite{riberi2025}
$$ \Delta \hat{b}(t)= 
\frac{ \nu^{-1/2} \,\Delta {\mathcal{O}}(t)}{ 
|\partial_b \langle {\mathcal{O}}(t) \rangle\vert_{b_0} |},$$
where $\Delta \mathcal{O}(t) \equiv [\langle \mathcal{O}(t)^2\rangle - \langle \mathcal{O}(t)\rangle^2]^{1/2}$, in terms of time-dependent noise-averaged expectation values, $ \langle \mathcal{O}(t) \rangle = \text{Tr}_\text{S} [ \bar{\rho}(t) \mathcal{O}]$, with $\bar{\rho}(t)\equiv \langle \rho(t) \rangle_\xi$.  The above estimation precision is lower-bounded by the quantum Cram\'er-Rao bound (QCRB),
$\Delta \hat b(t) \ge \big({ \nu\,F_Q(t) }\big)^{-1/2}
=\Delta \hat{b}_{\rm QCRB},$
where $F_Q(t)$ denotes the quantum Fisher information (QFI). One may write $F_Q(t)\equiv\mathrm{Tr} [\bar\rho(t) L^{\,2}]$, in terms of the symmetric logarithmic derivative (SLD) $L$ , determined by $\partial_b \bar\rho(t)=\tfrac12\{ L,\bar\rho(t)\}.$ The QCRB is saturable in the local regime when ${\mathcal{O}} \propto L$.

We focus on one-axis-twisted states (OATS), which are routinely accessible and robust to imperfect preparation. Starting from a coherent spin state (CSS) polarized along $\hat{\mathbf{z}}$, defined by $J_z|\mathrm{CSS}\rangle_{\hat{\mathbf{z}}}=J|\mathrm{CSS}\rangle_{\hat{\mathbf{z}}}$, such a state may be generated by first twisting about $\hat{\mathbf{x}}$, followed by a rotation about $\hat{\mathbf{z}}$:
\begin{equation}
|\mathrm{OATS}\rangle_{\hat {\bf z}}= e^{-i\beta J_z}\,e^{-i\mu  J_x^{\,2}}\,|\mathrm{CSS}\rangle_{\hat {\bf z}}.
\label{oats}
\end{equation}

\smallskip

\emph{Holstein-Primakoff mapping and Gaussian description.}
For analytic tractability, we work in the Holstein--Primakoff (HP) regime, where the state remains localized near the north pole of the Bloch sphere. Introducing a bosonic mode with $[a,a^\dagger]=1$ (in units $\hbar=1$) and dimensionless operator quadratures $\hat x=(a+a^\dagger)/\sqrt{2}$, $ \hat p=i(a^\dagger-a)/\sqrt{2}$, the HP mapping about $\hat {\bf z}$ \cite{HP} yields, to leading order,
$J_x \simeq \sqrt{J}\, \hat x,
 J_y \simeq \sqrt{J}\, \hat p,
 J_z =J-a^\dagger a.$
This approximation requires low excitations, $\langle a^\dagger a\rangle\ll 2J$. In particular, the twisting angle must satisfy $|\mu|=O(J^{-1/2})$, so that $\kappa\equiv J\mu=O(J^{1/2})$ \cite{SmerziRMP}.
Under the HP mapping, the Hamiltonian in Eq.\,\eqref{ham} becomes
\begin{equation}
H_{\mathrm{HP}}(t)=\sqrt{J}\big[b\, \hat x+\xi(t)\big( \hat x\cos\theta+ \hat p\sin\theta\big)\big],
\label{eq:HHP}
\end{equation}
which is linear in the quadratures. The OATS input in Eq.\,\eqref{oats} maps (up to a global phase) to the Gaussian state
\begin{equation}
|\Psi\rangle
= e^{+i\beta a^\dagger a}\,e^{-i\kappa \hat x^{\,2}}\,|0\rangle,
\label{eq:psiHP}
\end{equation}
where $|0\rangle$ is the vacuum state.

Because $H_{\rm HP}(t)$ is linear in $(\hat{x},\hat{p})$, the dynamics preserve Gaussianity. It is thus natural to describe quantum states via their Wigner function, $W(\mathbf r)$, with $\mathbf r\equiv (x,p)^\top$ being canonical phase space coordinates. A single-mode Gaussian state has the form
$$W(\mathbf r)=\frac{1}{2\pi\sqrt{\det\Sigma}}
\exp\!\left[-\frac12(\mathbf r-\boldsymbol\mu)^\top\Sigma^{-1}(\mathbf r-\boldsymbol\mu)\right],$$
where $\boldsymbol\mu$ and $\Sigma$ denote the mean vector and covariance  matrix. For the input state in Eq.\,\eqref{eq:psiHP}, the Wigner function is a centered Gaussian $W_0({\bf r})\equiv \mathcal N_{\mathbf r}(\boldsymbol\mu_0,\Sigma_0)$, with
\begin{equation}
\boldsymbol\mu_0=(0,0)^\top,\qquad
\Sigma_0=
\frac{1}{2}
\begin{pmatrix}
\delta & -2\eta\delta\\
-2\eta\delta & \delta^{-1}+4\eta^2\delta
\end{pmatrix},
\label{eq:Sigma0}
\end{equation}
in terms of parameters 
$\delta\equiv 1+2\kappa\sin2\beta+4\kappa^2\sin^2\beta$ and $
\eta\equiv \delta^{-1}\kappa (\cos2\beta+\kappa\sin2\beta).$ 
Here, $\delta$ sets the effective squeezing along $x$, while $\eta$ controls the shear between $x$ and $p$, as explicitly seen from
$$W_0(x,p)=\frac{1}{\pi}\exp\!\left[-\Big(\delta^{-1}x^2+\delta\,(p+2\eta x)^2\Big)\right].$$

\begin{figure*}[t!]
\centering
\includegraphics[width=18cm]{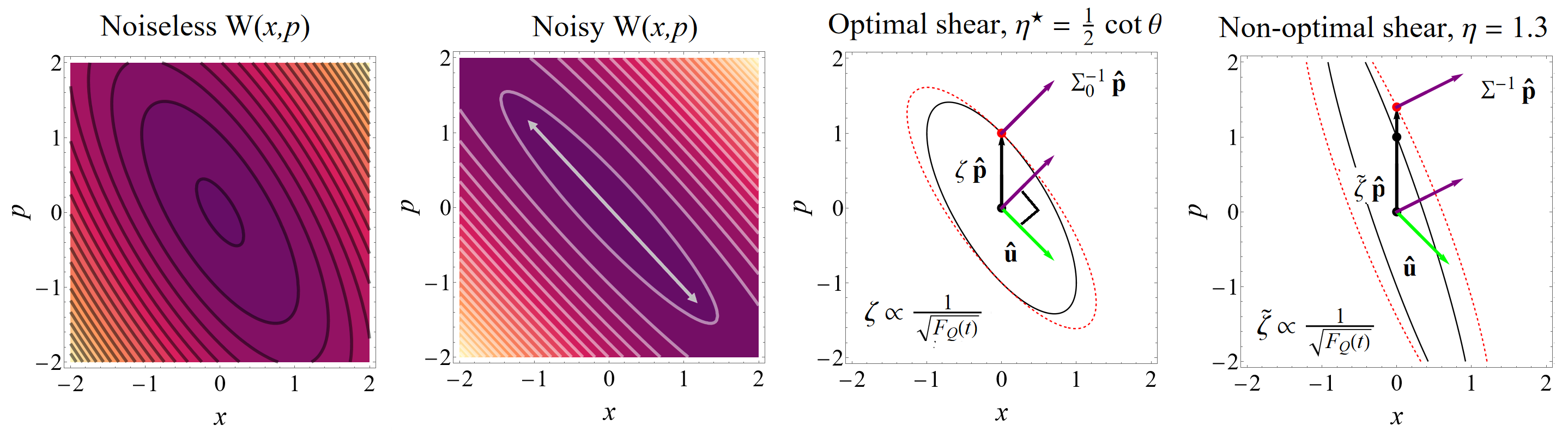}
\vspace*{-4mm}
\caption{{\bf 
Geometric protection against collective non-parallel directional noise.} 
Left: Gaussian Wigner function with mean and covariance given by Eq.\,(\ref{eq:muSigma_t}). Noise induces rank-one diffusion along $\hat{\mathbf u}$, broadening the ellipse  (white arrows). 
Right: Geometric interpretation of the QCRB. Contours $E(\mathbf{r}) \equiv \mathbf{r}^\top \Sigma_t^{-1} \mathbf{r}$ define the noiseless (black) and noisy (red, dashed) ellipses. The green arrow shows the direction $\mathbf{\hat{u}}$ of the rank-one contribution that modifies $\Sigma_0$ under noise, Eq.\,\eqref{eq:muSigma_t}. The purple arrows represent the vector $\Sigma^{-1} \mathbf{\hat{p}}$, which is normal to the noisy ellipse surface along $\hat{\mathbf{p}}$. For optimal shear ($\eta^\star= \frac{1}{2} \cot \tfrac{\pi}{4}  = \frac{1}{2} $ here, left), 
this equals the noiseless normal vector $\Sigma_0^{-1} {\bf \hat{p}}$. Translating this vector to the origin shows it fulfills 
Eq.\,(\ref{cond}). The noisy QCRB is set by the length of the black vector $\mathbf{r} = \zeta\, \mathbf{\hat{p}}$ joining the origin to the red surface ellipse along the displacement direction (red dot). At the noise-insensitive point (left), this contraction is unchanged by diffusion, restoring noiseless precision. This immunity fails for non-optimal shear (right), leading to a vector with larger magnitude $\tilde{\zeta}$ and worse performance. Parameters: $\theta= \tfrac{\pi}{4}$, $\delta=1$, $J=10$, $\gamma=2$, $t=0.6$. }
\label{Wigners}
\end{figure*}

\smallskip

\emph{Evolution and noise averaging.} For each fixed noise realization, the HP Hamiltonian in Eq.\,\eqref{eq:HHP} displaces the mean while leaving the covariance invariant. The conditional Wigner function is then 
$W_{\xi,t}(x,p)\equiv \mathcal N_{\mathbf r}\!\big(\boldsymbol\mu_{\xi}(t),\Sigma_0\big)$, with the updated mean given by
\begin{equation}
\boldsymbol\mu_\xi(t)
=-\sqrt{J}\,bt\,\hat{\mathbf p}
-\sqrt{J}\,Z(t)\,\hat{\mathbf u}.
\label{cW}
\end{equation}
Here, $\hat{\mathbf p}\equiv(0,1)^\top$ is the signal-imprint direction, $\hat{\mathbf v}= (\cos\theta,\sin\theta)^\top$ is the noise axis, and $\hat{\mathbf u}\equiv(-\sin\theta,\cos\theta)^\top$ is the corresponding diffusion direction, orthogonal to $\hat{\mathbf v}$.

Since $\xi(t)$ is Gaussian, the accumulated phase $Z(t)\equiv \int_0^t {\rm d}s\, \xi(s)$ is also zero-mean and Gaussian, with variance
$$\chi(t)=\langle Z(t)^2\rangle_\xi
=\int_0^t {\rm d}s\int_0^t {\rm d} s'\, C(s-s').$$
Averaging over $Z(t)$ removes the random displacement in Eq.\,\eqref{cW} and adds a {\em rank-one} diffusion contribution to the covariance in Eq.\,\eqref{eq:Sigma0}, 
\begin{equation}
\tilde{\boldsymbol\mu}(t)\equiv -\sqrt{J}\,b t\,\hat{\mathbf p},\qquad
\Sigma_t \equiv \Sigma_0+J\chi(t)\,\hat{\mathbf u}\hat{\mathbf u}^\top,
\label{eq:muSigma_t}
\end{equation}
yielding the Wigner function of the noise-averaged state, $\overline{W}({\bf r})=\mathcal{N}_{\bf r}(\tilde{\bm \mu}_t,\Sigma_t)$.
The corresponding equation of motion,
$\partial_t \overline{W}({\bf r})
=
\big[ \sqrt{J}\,b\,\partial_p
+ \tfrac 1 2 J \dot{\chi}(t)\,\partial_{\hat{\bf u}}^2\big]\,
\overline{W}({\bf r}),$
makes explicit that noise induces diffusion along $\hat{\mathbf u}$.

\smallskip

\emph{QFI and optimal readout.}
For a single-mode Gaussian state with parameter dependence only in the mean, the single-shot QFI and SLD are given by \cite{Haidong3,Pinel}:
\begin{equation}
F_Q(t)=\partial_b \boldsymbol\mu^\top \Sigma^{-1}\partial_b \boldsymbol\mu,
\qquad
L=\partial_b \boldsymbol\mu^\top \Sigma^{-1}(\hat{\mathbf r}-{\bm \mu}).
\label{eq:GaussianQFI_SLD}
\end{equation}
With $\nu=T/t$ repetitions, we simply have $F_Q^{\rm tot}(t)=\nu F_Q(t)$. In what follows, it is convenient to express the results in terms of the total number of sensors, $N=2J$. Substituting the covariance of Eq.\,\eqref{eq:muSigma_t} in Eq.\,(\ref{eq:GaussianQFI_SLD}) yields
\begin{equation}
F_Q^{\rm tot}(t)
=
\tfrac{1}{2} NT t\left(
\hat{\mathbf p}^\top\Sigma_0^{-1}\hat{\mathbf p}
-
\mathcal{C}(t)
\big|\hat{\mathbf p}^\top\Sigma_0^{-1}\hat{\mathbf u}\big|^2
\right),
\label{eq:Ftot}
\end{equation}
where $\mathcal{C}(t)\equiv {N\chi(t)}/[{2+N\chi(t)\,\hat{\mathbf u}^\top\Sigma_0^{-1}\hat{\mathbf u}}]>0$ is a noise-dependent coefficient. Thus, noise reduces the QFI {\em only} through the Fisher-metric overlap
$\hat{\mathbf p}^\top\Sigma_0^{-1}\hat{\mathbf u}$. If the input state is tuned such that
\begin{equation}
\hat{\mathbf p}^\top\Sigma_0^{-1}\hat{\mathbf u}
= 2\delta (\cos \theta - 2\eta \sin\theta) = 0, 
\label{cond}
\end{equation}
the noise-induced degradation term vanishes. This corresponds to an initial optimal OAT for which the shear satisfies $\eta^\star=\tfrac{1}{2}\cot\theta$.
Although diffusion continues to broaden the state, its contribution to the QFI vanishes, leaving the estimation precision identical to the noiseless value. From Eq.\,\eqref{eq:GaussianQFI_SLD}, the optimal measurement is the quadrature $Q_{\rm opt}\propto \mathbf q_{\rm opt}^\top {\mathbf r}$, with
$\mathbf q_{\rm opt}=\Sigma_t^{-1}\hat{\mathbf p}$.
At the noise-insensitive point, $\mathbf q_{\rm opt}$ is parallel to $\hat{\mathbf v}$, yielding the optimal quadrature $Q_{\rm opt}\propto \cos\theta\, \hat x+\sin\theta\, \hat p.$ In terms of the original spin variables, this corresponds to measuring along the noise axis,
$J_{\hat{\mathbf v}}=\cos\theta\,J_x+\sin\theta\, J_y.$
Geometrically, as seen in Fig.\,\ref{Wigners}, diffusion elongates the Wigner ellipse along $\hat{\mathbf{u}}$, but at the noise-insensitive point $\eta^\star$, the inverse-covariance contraction along $\hat{\mathbf{p}}$ remains unchanged. Thus, the curvature along the signal direction is preserved despite noise-induced broadening. 

\begin{figure*}[t!]
\centering
\includegraphics[width=16cm]{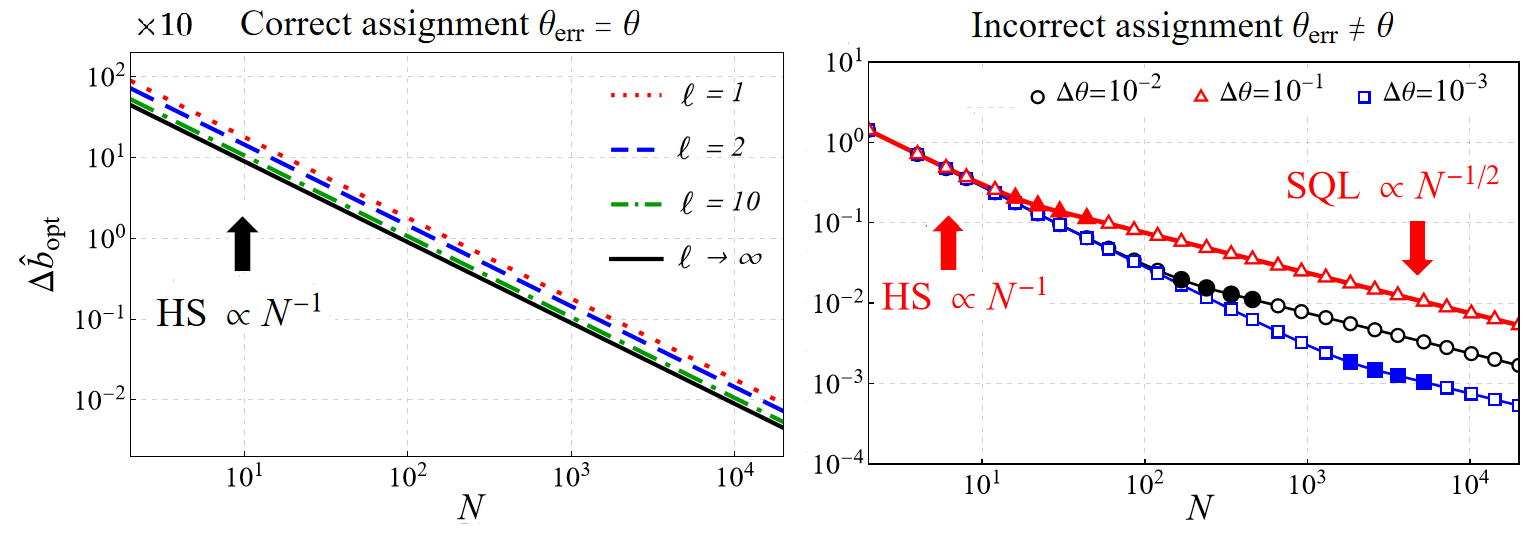}
 \vspace*{-4mm}
\caption{{\bf Optimal performance of OATS under collective directional noise.} 
Left: Lower bound to precision $\Delta \hat{b}_{\rm opt}= [F_{Q, \star}^{\rm tot}(t^\star)]^{-1/2}$ for different values of the short-time decay coefficient $\chi(t) \simeq \chi_0^{\ell} (\omega_c t)^\ell$ as a function of probe number $N$. Full knowledge of the noise direction is assumed. In all cases, HS is achieved, $\Delta \hat{b}_{\rm opt} \propto N^{-1}$, with the largest constant prefactor for Markovian noise ($\ell=1$) and increasingly better performance with higher $\ell$. Right: Transition from HS to noise-limited asymptotics when measuring a suboptimal quadrature ${Q}_{\rm err}= \cos \theta_{\rm err}\,\hat x + \sin \theta_{\rm err} \,\hat{p}$. Here, imperfect knowledge of the noise axis leads to an incorrect assignment $\theta_{\rm err}$ that deviates from the true noise direction $\theta$. For temporally correlated noise, $\ell=2$, the performance is ultimately bounded by the SQL, with the regime change -- represented by solid plot markers -- occurring at higher probe number as $\Delta \theta$ decreases.  In both figures, we used $\chi_0 =1$, 
$\theta =\pi/4$, 
and $\omega_c=1$. 
}
\label{QCRB} 
\end{figure*}

\smallskip

\emph{Heisenberg scaling under arbitrary temporal correlations.}
At the optimal shear $\eta^\star(\theta)$, Eq.~\eqref{eq:Ftot} for the QFI simply becomes
$F_{Q,\star}^{\rm tot}(t)= N T t\,\delta.$
Maximizing this expression reduces to optimizing $\delta$ and $t$, subject to the HP validity constraint. For OATS inputs, the feasible parameters may be shown to satisfy 
$$\tfrac{1}{4}(1+2\kappa^2-\delta)^2+(\eta\delta)^2=\kappa^2(1+\kappa^2),$$
from which the condition $\delta \geq \delta_{\rm min}= \sin \theta$ follows for $\eta^{\star}$ (see also Appendix A in \cite{Suppl}). We may enforce the validity of the HP approximation by imposing an excitation cap $0< \lambda \ll 1$,
$$\langle a^\dagger a\rangle(t)
=
\tfrac{1}{2}\big(\mathrm{Tr}\,\Sigma_0-1\big)
+\tfrac{1}{4}N\chi(t)
\le \tfrac{1}{2}\lambda N.$$
 
Assume that the decay obeys a short-time scaling of the form $\chi(t)\simeq \chi_0^{\,\ell}(\omega_c t)^{\ell}$, with $\ell=1$ (Markovian) or $\ell>1$ (non-Markovian), with the frequency parameter $\omega_c$ being identifiable with an inverse correlation time in the latter case \cite{third}. We can then optimize by saturating the excitation cap at $t=t^\star$. Decomposing the excitation budget into a contribution $\alpha$ for noise-induced broadening, and a contribution $\lambda/2-\alpha$ corresponding to the input, 
\begin{align}
\mathrm{Tr}\,\Sigma_0-1= (\lambda-2\alpha)N,\qquad
\tfrac{N}{4}\chi(t^\star)=\alpha N,
\label{exc}
\end{align}
yields the encoding time $t^\star=(4\alpha)^{1/\ell}/(\chi_0\omega_c)$. 
Maximizing over the fraction  $\alpha$ gives in turn $\alpha^\star=\lambda/[2(\ell+1)]$. Using $\mathrm{Tr}\,\Sigma_0=\tfrac{1}{2}\big(\delta/\sin^2\theta+\delta^{-1}\big)
\simeq\delta/(2\sin^2\theta)$, valid in the regime $\delta \simeq \mathcal{O}(N)$, together with the excitation constraint in Eq.\,(\ref{exc}), leads to the condition $\delta(\alpha)\simeq2(\lambda-2\alpha)N\sin^2\theta$. From this, $\delta^\star=2\ell\lambda N\sin^2\theta/(\ell+1)$ follows upon replacing $\alpha^\star$, which determines the optimal initial OATS parameters $(\beta^\star, \mu^\star)$ (see Appendix A in \cite{Suppl}). Note that since the quadrature squeezing must obey $\delta^\star \geq \delta_{\rm min}$, there is a minimal dephasing angle $\theta \geq \theta_{\rm min} \simeq (\lambda N)^{-1} [(\ell+1)/2\ell]$, beyond which the procedure is not applicable. In other words, as the noise approaches the pure-dephasing limit, the geometric protection mechanism breaks down, consistent with intuition. Substituting $t^\star$ and $\delta^\star$ into $F_{Q,\star}^{\rm tot}(t)$, 
we finally obtain
\begin{equation}
F_{Q,\star}^{\rm tot}(t^\star)
\simeq
\left[
\frac{\sin^2\theta}{\chi_0}
\left(\frac{2\lambda}{\ell+1}\right)^{\!\frac{1}{\ell}}
\frac{2\ell\lambda}{\ell+1}
\right]
\frac{T}{\omega_c}\,N^2.
\label{eq:Fstar_final}
\end{equation}
Thus, $\Delta \hat b_{\rm opt}\propto N^{-1}$, with HS being achieved for any $\ell$.

\smallskip

\emph{Sensitivity analysis.} We finally address the extent to which the HS is robust against imperfect knowledge of the noise axis, $\theta$. Designing the probe shear for QFI protection at an erroneous angle $\theta_{\rm err}$ which deviates from the actual value by $\Delta\theta \equiv \theta - \theta_{\rm err}$, yields a precision
\begin{equation}
\Delta \hat b(t)=\frac{1}{\sqrt{NT t}}
\left[
\frac{1}{\delta}
+N\chi(t) \frac{ \sin^2 \Delta \theta}{ \sin^2\theta_{\rm err}}
\right]^{1/2}.
\label{eq:delta_b_mismatch}
\end{equation}
The second term vanishes for $\Delta \theta=0$, but otherwise reintroduces a dependence upon the noise spectral properties.  As $N$ increases, this produces a crossover from HS to collective-dephasing behavior, with the asymptotics determined by the nature of the temporal correlations -- in particular, SQL-limited precision for stationary colored noise  (Fig.~\ref{QCRB}). Further including a small $\hat {\bf z}$ noise component enriches the analysis but does not qualitatively alter the main conclusions (see Appendix B in \cite{Suppl}). 

\smallskip

\emph{Conclusion.} In this work, we have identified a geometric mechanism that enables Heisenberg-limited frequency estimation in the presence of \emph{collective, non-parallel directional noise} with arbitrary temporal correlations. Working in the HP regime, we can tailor the shear of an initial OATS so that the signal-induced displacement becomes orthogonal to the noise-induced diffusion direction in the Fisher metric defined by the initial covariance. As a result, the QFI becomes insensitive to noise-induced diffusion for an optimal selection of the shear parameter. Heisenberg-limited performance is then achieved using a fixed, non-adaptive measurement of a single collective spin component, without resorting to quantum error correction, external control, or ancillary degrees of freedom.

Our mechanism differs from signal protection via decoherence-free subspaces which, as noted, have also been explored as a strategy to counter collective noise \cite{Dorner2012,Landini2}. Here, the noise acts non-trivially on the state and generally reduces its purity but, thanks to the directionality properties, a specific estimation functional is nonetheless protected. In this respect, our mechanism is closer in spirit to a noiseless subsystem, whereby a degree of freedom (classical, in this case) is preserved despite the state being degraded \cite{klm2000,evan2003}.
Operationally, if the noise assumptions are met, the proposed protocol requires only the preparation of suitably rotated OATS inputs and collective spin readout, both of which are available in current atomic interferometry platforms \cite{riedel2010,BEC2,Colombo}.


Future work will investigate the usefulness of the geometric noise-immunity mechanism in other collective-noise scenarios of interest. For instance, settings where noise cannot be modeled classically and additional quantum contributions arise \cite{FelixPRA}, or sequential protocols where metrological scaling advantage is sought in time \cite{Giovannetti0}, are both relevant avenues worth exploring.  We will also study the interplay between collective directional noise and continuous driving, beyond the HP Gaussian regime. Understanding the robustness of the approach to state preparation and measurement (SPAM) errors is also an important step toward experimental implementation, along with the possibility of designing SPAM-mitigated protocols \cite{khan2024}. Finally, an interesting  
direction is to explore whether noise engineering techniques can be harnessed to enforce a directional structure beneficial to metrological scaling.


\smallskip

\emph{Acknowledgments.}
This work is supported by DOE EXPRESS award No.\,DE-SC0024685. Additional support by the DOE award No.\,DE-SC0026373 is acknowledged.

\bibliography{NoisyRamseyBib}

\end{document}


\title{{\large Supplemental \vspace*{.5mm}Material for:} \\ 
Heisenberg scaling under collective non-parallel directional noise via geometric state design}

\author{Francisco Riberi}
\affiliation{\mbox{Electrical and Computer Engineering Department, University of New Mexico, 498 Terrace St NE, Albuquerque, NM 87106}} 

\author{Lorenza Viola}
\affiliation{\mbox{Department of Physics and Astronomy, Dartmouth College, 6127 Wilder Laboratory, Hanover, New Hampshire 03755, USA}} 

\author{Milad Marvian} 
\affiliation{Electrical and Computer Engineering Department, University of New Mexico, 498 Terrace St NE, Albuquerque, NM 87106}

\date{\today}

\maketitle

\vspace*{-10mm}


\setcounter{figure}{0}
\setcounter{equation}{0}
\setcounter{section}{0}

\appendix
\section{Mapping optimal Gaussian parameters to OATS variables}
\label{app:OATSmap}

Here, we show how to construct OAT input states that realize the optimal Gaussian parameters $\{\delta^\star,\eta^\star\}$ identified in the main text.
The problem amounts to inverting the relations between the OATS parameters
$\{\kappa,\beta\}$ (equivalently, $\{\mu,\beta\}$ with $\kappa=J\mu$)
and the Gaussian covariance parameters $\{\delta,\eta\}$ appearing in the initial covariance matrix $\Sigma_0$.
 Starting from
\begin{align}
\delta
&=
1+2\kappa\sin2\beta
+4\kappa^2\sin^2\beta, \qquad
\eta
= \delta^{-1}
\kappa(\cos2\beta+\kappa\sin2\beta), \label{deltaeta}
\end{align}
and using
\(
4\sin^2\beta=2(1-\cos2\beta),\)
the first equation becomes
\[
1+2\kappa^2-\delta
=
2\kappa(\kappa\cos2\beta-\sin2\beta).
\]
Squaring this expression and adding $(\eta\delta)^2$ gives
\begin{align}
\frac14(1+2\kappa^2-\delta)^2+(\eta\delta)^2
&=
\kappa^2
\Big[
(\kappa\cos2\beta-\sin2\beta)^2
+
(\cos2\beta+\kappa\sin2\beta)^2
\Big]=
\kappa^2(1+\kappa^2).
\label{identity}
\end{align}
This identity constrains the Gaussian parameters generated by OATS and allows $\kappa$ to be directly obtained from $(\delta,\eta)$. Substituting the optimal Gaussian parameters,
\begin{align}
\eta^\star=\tfrac12\cot\theta, \qquad \delta^\star
=
\frac{4J\ell\lambda\sin^2\theta}{\ell+1}, \label{etadeltaopt}
\end{align}
 and solving for $\kappa^2$ from Eq.\,(\ref{identity}) gives
\begin{equation}
\kappa^{\star2}
=
\frac{(1+4\eta^{\star2})\,\delta^{\star2}-2\delta^\star+1}
{4\delta^\star}.
\end{equation}
Introducing the auxiliary variable
$R\equiv \delta^\star-1-2\kappa^{\star2},$ the two relations in Eq.\,\eqref{deltaeta} may then be rewritten as
\begin{align}
\sin2\beta^\star-\kappa^\star\cos2\beta^\star
&=
\frac{R}{2\kappa^\star}, \qquad
\kappa^\star\sin2\beta^\star+\cos2\beta^\star
=
\frac{\eta^\star\delta^\star}{\kappa^\star}.
\end{align}
Solving this linear system gives
\begin{align}
\sin2\beta^\star
&=
\frac{
R+2\kappa^\star\eta^\star\delta^\star
}{
2\kappa^\star(1+\kappa^{\star2})
}, \qquad
\cos2\beta^\star
=
\frac{
2\eta^\star\delta^\star-\kappa^\star R
}{
2\kappa^\star(1+\kappa^{\star2})
}.
\end{align}
Consequently, the optimal rotation angle satisfies:
\begin{equation}
\beta^\star
=
\frac12
\arctan\!\left(\frac{
R+2\kappa^\star\eta^\star\delta^\star}
{2\eta^\star\delta^\star-\kappa^\star R}
\right).
\end{equation}
Mapping back to the spin system, the optimal squeezing angle is $\mu^\star=\kappa^\star/J$. This satisfies $\mu^\star = O(J^{-1/2})$,
ensuring the validity of the HP approximation for all noise models we considered.

At the optimal shear $\eta^\star=\frac12\cot\theta,$ we have
$
1+4\eta^{\star2}=\csc^2\theta.$
Equation~\eqref{identity} then becomes
\[
\delta^{\star 2}
-
2(1+2\kappa^{\star 2})\sin^2\theta\,\delta^\star
+
\sin^2\theta
=
0,
\]
whose solutions are
\[
\delta^\star
=
(1+2\kappa^{\star 2})\sin^2\theta
\pm
\sin\theta
\sqrt{(1+2\kappa^{\star 2})^2\sin^2\theta-1}.
\]
The positive branch corresponds to the metrologically relevant solution, as the protected QFI is proportional to $\delta$. Since the square root is real only when
\(
(1+2\kappa^{\star 2})\sin\theta\ge1,\)
we have the inequality
\(
\delta^\star
\ge
(1+2\kappa^{\star 2})\sin^2\theta
\ge
\sin\theta,\) whereby the lower bound stated in the main text, 
\(
\delta^\star\ge\delta_{\min}=\sin\theta
\),
follows.

\section{Precision from quadrature measurements with imperfect knowledge of the noise direction}
\label{app:quadrature_precision}

Here, we show how the estimation precision can be computed explicitly when the noise direction is not perfectly known, and clarify when and why the noise spectrum re-enters the precision bound.
As described in the main text, we work in the Gaussian regime, where the noise-averaged state is fully characterized by a mean vector and covariance matrix. 
The noise-averaged Wigner function takes the form
\begin{equation}
W(\boldsymbol{r}) \propto 
\exp\!\left[
-\tfrac12(\boldsymbol{r}-\boldsymbol{\mu})^{\mathsf T}
\Sigma^{-1}
(\boldsymbol{r}-\boldsymbol{\mu})
\right],
\end{equation}
with phase-space coordinates $\boldsymbol{r}=(x,p)^{\mathsf T}$ and mean vector and covariance matrix given by
 \begin{equation}
\boldsymbol{\mu}(b)=-\sqrt{J}\,b t\,{\bf\hat{p}}, \qquad \Sigma_t(\theta)=\Sigma_0 + J\chi(t)\,{\bf\hat{u}}{\bf\hat{u}}^{\mathsf T}, \qquad  \Sigma_0=
\frac12
\begin{pmatrix}
\delta &-2\eta\delta\\
-2\eta\delta&
\delta^{-1}+4\eta^2\delta
\end{pmatrix}.
 \label{meancov}
 \end{equation}
Here, $\delta$ $\eta$ are input-state parameters, $\chi(t)$ is the noise-induced diffusion coefficient and ${\bf\hat{u}}=(-\sin\theta,\cos\theta)^{\mathsf T}$ specifies the direction along which diffusion occurs.
To estimate the target parameter $b$, we perform homodyne detection of a general quadrature
\begin{equation}
 Q_\phi = \cos\phi\,\hat {x} + \sin\phi\,\hat {p}
= \hat{\bf v}^{\mathsf T} (\hat x,\hat p),
\qquad
\hat{\bf v}=(\cos\phi,\sin\phi).
\end{equation}
We compute the squared estimation error associated with measuring $ Q_\phi$ over a total interrogation time $T$  using error propagation, with the relevant expectation values written in terms of the mean and covariance in Eq.\,(\ref{meancov}):
\begin{equation}
\Delta^2\, \hat b[ Q_\phi](t)
=
\frac{t\,\Delta^2 Q_\phi}
{T\,(\partial_b\langle \hat Q_\phi\rangle)^2}, \qquad \langle Q_\phi \rangle
=
\hat{\bf v}^{\mathsf T}\boldsymbol{\mu},
\qquad
\Delta^2 Q_\phi
=
\hat{ \bf v}^{\mathsf T}\Sigma\,\hat{\bf v}.
\label{eq:quadrature_error}
\end{equation}

If the noise direction $\theta$ is known exactly, the best strategy consists
of choosing the input shear parameter
$\eta^\star=\tfrac12\cot\theta$ and measuring the quadrature
$\phi=\theta$, as shown in the main text.
The measured quadrature direction
\(
\hat{\mathbf v}
=(\cos\theta,\sin\theta)^{\mathsf T}\)
is orthogonal to the diffusion direction
\(
\hat{\mathbf u}
=(-\sin\theta,\cos\theta)^{\mathsf T},\)
so that
\(
\hat{\mathbf v}^{\mathsf T}\hat{\mathbf u}=0.\)
The optimal shear simultaneously satisfies the Fisher-metric orthogonality
condition
\(
\hat{\mathbf p}^{\mathsf T}\Sigma_0^{-1}\hat{\mathbf u}=0,
\)
which is precisely the condition for the QFI to become independent of the
noise-induced diffusion.
Consequently, the diffusion contribution to the measured quadrature variance
vanishes and the estimation precision becomes independent of $\chi(t)$.
Substituting into Eq.\,\eqref{eq:quadrature_error}, with $N=2J$, one finds
$\Delta^2\,\hat b(t)
=
{1}/{(N\delta Tt)}
=
(\Delta\hat b_{\rm QCRB})^2,$
which saturates the QCRB, independently of the noise spectrum.

\subsection{Effect of an in-plane directional error}

Assume first that the true noise direction $\theta$ is not precisely known, but the noise is known to act in the $xy$ plane ($v_z=0$), as assumed in Eq.\,(1) in the main text. The experimenter may then fix a measurement quadrature $\phi=\theta_{\mathrm{err}}$ and tune the input shear accordingly, while the actual diffusion direction differs by $\Delta \theta=\theta-\theta_{\mathrm{err}}$. A direct evaluation of the quadrature moments yields
\begin{equation}
\langle  Q_{\theta_{\mathrm{err}}}\rangle
=
-\sqrt{J}\,b t\,\sin\theta_{\mathrm{err}}, \quad
\Delta^2 Q_{\theta_{\rm err}}
=
\frac{\sin^2\theta_{\rm err}}{2\delta}
+
J\chi(t)\sin^2\Delta\theta.
\end{equation}
Substituting into Eq.\,\eqref{eq:quadrature_error}, again with $N=2J$, gives the precision reported in Eq.\,(13) of the main text, namely, 
\begin{equation}
\Delta^2 \hat b(t)
=
\frac{1}{N T t}
\left[
\frac{1}{\delta}
+
N\chi(t)\frac{\sin^2 \Delta \theta}
{\sin^2\theta_{\mathrm{err}}}
\right].
\label{eq:quadrature_mismatch}
\end{equation}
Equation~\eqref{eq:quadrature_mismatch} makes the physical mechanism transparent. When the assumed noise direction matches the true one ($\Delta \theta=0$), the protocol remains noise-immune and saturates the QCRB. When $\Delta\theta \neq0$, diffusion leaks into the measured quadrature, introducing an additive noise term proportional to $N \chi(t)$. In this regime, the attainable precision becomes sensitive to the short-time behavior of $\chi(t)$, and hence to the noise spectrum.

Quadrature measurements therefore provide an operational way to interpolate between geometry-protected metrology and noise-limited sensing when full knowledge of the noise direction is unavailable.

\subsection{Effect of an out-of-plane directional error}
\label{app:out_of_plane_error}
We now consider an input state with the same shear
$\eta=\tfrac12\cot\theta_{\rm err}$ and an arbitrary Gaussian squeezing
parameter $\delta>0$, but allow the true collective-noise direction to have a small component $v_z$ 
along the polarization axis, that is, 
\begin{equation}
\hat{\mathbf v}
=
\left(
\sqrt{1-\epsilon^2}\cos\theta,\,
\sqrt{1-\epsilon^2}\sin\theta,\,
\epsilon
\right).
\end{equation} 
In the HP approximation, and up to an irrelevant stochastic scalar, the Hamiltonian becomes
\begin{equation}
H_{\rm HP}(t)
=
\sqrt{J}\,b\hat x
+
\sqrt{J(1-\epsilon^2)}\,\xi(t)
\left(
\cos\theta\,\hat x+\sin\theta\,\hat p
\right)
-
\epsilon\, \xi(t)a^\dagger a .
\label{eq:HP_out_of_plane}
\end{equation}
The Heisenberg equations remain linear in the quadrature operators, so the dynamics consist of a stochastic phase-space rotation and displacement. Consequently, while the noise-averaged state becomes non-Gaussian due to the presence of the quadratic term, the first and second moments of any quadrature can still be evaluated exactly by Gaussian averaging over the noise realizations. Here, we consider readout $Q_{\theta_{\rm err}}$ designed using the estimate $\theta_{\rm err}$ for $v_z=0$.

Defining the auxiliary variables 
\[
X_{\epsilon}(t) =\epsilon^2\chi(t),
\qquad
{\cal A}_{\epsilon}(t)
\equiv 
\int_0^t{\rm d}\tau\,
e^{-\epsilon^2\chi(\tau)/2},
\qquad {\cal D}_{\epsilon,\delta}(t)
\equiv
\frac{e^{-2X_{\epsilon}(t)}}{\delta}
+
\frac{1-e^{-2X_{\epsilon}(t)}}
{2\sin^2\theta_{\rm err}}
\left[
\frac{\delta}{\sin^2\theta_{\rm err}}
+
\frac1\delta
\right]
,
\]
and
\begin{align}
{\cal C}_{\epsilon,\Delta\theta}(t)
\equiv {}&
\frac{1-\epsilon^2}
{2\epsilon^2\sin^2\theta_{\rm err}}
\left[
1-e^{-X_{\epsilon}(t)}
+
\left(
e^{-2X_{\epsilon}(t)}-e^{-X_{\epsilon}(t)}
\right)
\cos(2\Delta\theta)
\right],
\label{eq:C-epsilon-deltatheta}
\end{align}
the relevant sensitivity and variance can be written as
\begin{equation}
\left.
\partial_b
\langle Q_{\theta_{\rm err}}(t)\rangle
\right|_{b=0}
=
-\sqrt{N/2}\,
{\cal A}_\epsilon(t)
\sin\theta_{\rm err}, \qquad
 \;\Delta^2 Q_{\theta_{\rm err}} 
=
\frac{\sin^2\theta_{\rm err}}{2}
\left[
{\cal D}_{\epsilon,\delta}(t)
+
N{\cal C}_{\epsilon,\Delta\theta}(t)
\right].
\label{eq:Qerr-moments}
\end{equation}
Using error propagation, the resulting uncertainty is
\begin{equation}
\Delta^2\,\hat b(t)
=
\frac{t}{T{\cal A}_{\epsilon}(t)^2}
\left[
\frac{{\cal D}_{\epsilon,\delta}(t)}{N}
+
{\cal C}_{\epsilon,\Delta\theta}(t)
\right].
\label{eq:out-of-plane-arbitrary-delta}
\end{equation}
Here, ${\cal D}_{\epsilon,\delta}(t)$ contains the intrinsic squeezed
variance together with the additional variance generated by the stochastic
rotation of the input covariance, whereas
${\cal C}_{\epsilon,\Delta\theta}(t)$ is the additive contribution generated
by the stochastic displacement. In the perfectly aligned limit
$\epsilon=\Delta\theta=0$, one has
\[
{\cal D}_{0,\delta}(t)=\frac1\delta,
\qquad
\lim_{\epsilon\to0}{\cal C}_{\epsilon,0}(t)=0,
\]
and the 
result corresponding to known directionality is again recovered.

To discuss the probe-number scaling under directional errors, we now specialize to a family of
Gaussian input states with a linearly increasing squeezing parameter as in Eq. (\ref{etadeltaopt}),
\(
\delta=\zeta N,\)
for \(
\zeta>0,\)
independent of $N$. 
Equation~\eqref{eq:out-of-plane-arbitrary-delta}
then becomes
\begin{equation}
\Delta^2\,\hat b(t)
=
\frac{t}{T{\cal A}_{\epsilon}(t)^2}
\left[
\frac{{\cal D}_{\epsilon}(t)}{N^2}
+
{\cal C}^{\rm tot}_{\epsilon,\Delta\theta}(t)
\right],
\label{eq:out-of-plane-specialized}
\end{equation}
where
\begin{align}
{\cal D}_{\epsilon}(t)
\equiv{}
\frac{e^{-2X_{\epsilon}(t)}}{\zeta}
+
\frac{1-e^{-2X_{\epsilon}(t)}}
{2\zeta\sin^2\theta_{\rm err}},
\qquad
{\cal C}^{\rm tot}_{\epsilon,\Delta\theta}(t)
\equiv {}
\frac{\zeta}
{2\sin^4\theta_{\rm err}}
\left[
1-e^{-2X_{\epsilon}(t)}
\right]
+
{\cal C}_{\epsilon,\Delta\theta}(t).
\end{align}
Thus, at fixed interrogation time, the uncertainty separates into a
Heisenberg-limited contribution proportional to $N^{-2}$ and a
directional-error contribution determined by
${\cal C}^{\rm tot}_{\epsilon,\Delta\theta}(t)$. After optimizing the
encoding time, the large-$N$ precision becomes determined by the short-time
behavior of the noise correlations; in particular, for $\ell=2$ the
asymptotic scaling is SQL. Consequently,
\(
N^2{\cal C}^{\rm tot}_{\epsilon,\Delta\theta}(t^\star)
\ll
{\cal D}_{\epsilon}(t^\star)
\)
is the condition under which HS
is preserved. For small $z$-directional errors, $X_{\epsilon}(t)=\epsilon^2\chi(t)\ll1$, the total
error coefficient has the expansion
\begin{align}
{\cal C}^{\rm tot}_{\epsilon,\Delta\theta}(t)
={}&
\chi(t)
\left[
\frac{(\Delta\theta)^2}{\sin^2\theta_{\rm err}}
+
\frac{\zeta\epsilon^2}{\sin^4\theta_{\rm err}}
\right]+
O\Big(
\epsilon^2\chi(t)(\Delta\theta)^2,\,
\epsilon^2\chi(t)^2,\,
\chi(t)(\Delta\theta)^4
\Big).
\label{eq:Ctot-small-errors}
\end{align}
The crossover out of the HS regime occurs when the two leading-order terms in Eq.\,\eqref{eq:out-of-plane-specialized} become comparable, that is, we may write 
\(
\frac{{\cal D}_{\epsilon}(t^\star)}{N_{\rm cross}^2}
\sim
{\cal C}^{\rm tot}_{\epsilon,\Delta\theta}(t^\star)\), whereby
\begin{equation}
N_{\rm cross}
\sim \left(
\frac{{\cal D}_{\epsilon}(t^\star)}
{\chi(t^\star)}\right)^{1/2} \left(
\dfrac{\Delta\theta^2}{\sin^2\theta_{\rm err}}
+
\dfrac{\zeta\epsilon^2}{\sin^4\theta_{\rm err}}
\right)
^{-1/2}.
\label{eq:Ncross-out-of-plane}
\end{equation}
Up to the geometry-dependent factors and the smooth coefficient
${\cal D}_{\epsilon}(t^\star)$, this gives
\[
N_{\rm cross}
\propto
\left[
(\Delta\theta)^2+\epsilon^2
\right]^{-1/2}.
\]
Therefore, the in-plane angular error $\Delta\theta$ and the out-of-plane error $\epsilon$ enter the $N$-crossover at the same perturbative order: both generate quadratic directional-error contributions -- through $(\Delta\theta)^2$ and $\epsilon^2$, respectively -- and therefore reduce the extent of the HS regime at the same parametric order.
